\documentclass[12pt]{iopart}
\usepackage{iopams}

\input epsf.sty

\newlength{\TZ}
\newcommand{\BEQ}{\begin{equation}}     
\newcommand{\BEA}{\begin{eqnarray}}
\newcommand{\EEQ}{\end{equation}}       
\newcommand{\EEA}{\end{eqnarray}}
\newcommand{\D}{{\rm d}}                
\newcommand{\II}{{\rm i}}               
\newcommand{\wit}[1]{\widetilde{#1}}    

\renewcommand{\vec}[1]{\boldsymbol{#1}} 

\newcommand{\appsection}[2]{\setcounter{equation}{0}\setcounter{subsection}{0}
\section*{Appendix #1. #2}
\renewcommand{\theequation}{#1\arabic{equation}}
              \renewcommand{\thesection}{#1} }

\catcode`\@=11
\def\numberbysection{\@addtoreset{equation}{section}
        \def\theequation{\thesection.\arabic{equation}}}

\begin{document}
\title[Semi-infinite kinetic spherical model]{Out-of-equilibrium properties of the semi-infinite
kinetic spherical model}

\author{Florian Baumann$^{1,2}$ and Michel Pleimling$^1$}
\address{$^1$Institut f\"ur Theoretische Physik I, 
Universit\"at Erlangen-N\"urnberg, \\
Staudtstra{\ss}e 7B3, D -- 91058 Erlangen, Germany}
\address{$^2$Laboratoire de Physique des 
Mat\'eriaux,\footnote{Laboratoire associ\'e au CNRS UMR 7556.} 
Universit\'e Henri Poincar\'e Nancy I, \\ 
B.P. 239, F -- 54506 Vand{\oe}uvre l\`es Nancy Cedex, France}

\begin{abstract}
We study the ageing properties of the semi-infinite kinetic spherical model at the critical point and
in the ordered low-temperature phase, both for Dirichlet and Neumann boundary conditions. The 
surface fluctuation-dissipation ratio and the scaling functions
of two-time surface correlation and response functions are determined explicitly in the dynamical scaling regime.
In the low-temperature phase our results show that for the case of Dirichlet boundary conditions the value of the non-equilibrium
surface exponent $b_1$ differs from the usual bulk value of systems undergoing phase ordering.
\end{abstract}

\section{Introduction}
Ageing phenomena encountered in systems with slow degrees of freedom are due to relaxation 
processes that depend on the thermal history of the sample. Ageing is found in various systems,
ranging from structural glasses and spin glasses to colloids and polymers \cite{Stru78,Cate00,Cugl03}. In additon, ageing phenomena
are also observed in ferromagnets quenched to or below their critical point \cite{Bray94,Godreche2002,Henk04,CalGam05}.

Ageing processes are best revealed through the study of two-time quantities like dynamical correlation and
response functions. Well-known examples are given by the autocorrelation and autoresponse functions
\BEQ
C(t,s) = \left\langle \phi(t) \phi(s)\right\rangle \;\; , \;\;
R(t,s) = \left.\frac{\delta \langle\phi(t)\rangle}{\delta h(s)}\right|_{h=0} ~~(t > s)
\EEQ
where $\phi(t)$ is the time-dependent order parameter and $h$ is the field conjugate to $\phi$. The
time $t$ elapsed since the quench is usually called observation time and $s$ is the waiting time.
In out-of-equilibrium systems one often observes in the regime $t$, $s$, $t-s \gg t_{micro}$ ($t_{micro}$ 
being a microscopical time scale) the following dynamical scaling behaviour \cite{Cugl03}:
\BEQ 
C(t,s) = s^{-b} f_{C}(t/s) \;\; , \;\;
R(t,s) = s^{-1-a} f_{R}(t/s).
\EEQ
Here $a$ and $b$ are non-equilibrium exponents, whereas $f_{C}(y)$ and $f_R(y)$ are scaling functions 
which for $y \gg 1$ show the power-law behaviours
\BEQ 
f_{C}(y) \sim y^{-\lambda_C/z} \;\; , \;\;
f_{R}(y)\sim y^{-\lambda_R/z}
\EEQ
where $\lambda_C$ and $\lambda_R$ are called autocorrelation \cite{Fish88,Huse89} and autoresponse exponents
\cite{Pico02}, and $z$ is the dynamical
exponent. These simple scaling forms are observed in systems which are characterised by the existence of a time-dependent
length scale $\xi \sim t^{1/z}$.

Specifically, for ferromagnets we distinguish between physically very different situations. Below $T_c$ phase 
ordering takes place, and the slow degrees of freedom are provided by the moving interfaces separating different
ordered domains. We then commonly have $b=0$, whereas $a$ takes on the value $1/z$ or $(d-2 + \eta)/z$, depending on whether
the static correlations decay exponentially or whether they follow
a power-law with an exponent $d-2 +\eta$ \cite{Henk03,Henk04a},
where $d$ is the number of space dimensions. At the critical point general scaling arguments lead to the relation
$a = b = (d-2 + \eta)/z$ where $\eta$ is the well-known static critical exponent. Furthermore, the autocorrelation
exponent $\lambda_C$ can be related to the initial slip exponent \cite{Janssen89} describing the relaxation of the
order parameter in the short-time regime \cite{Jans92}. For initial states with short-range correlations one finds $\lambda_C  =
\lambda_R$, but this relation may be broken when starting from long-range correlated initial states \cite{Pico02}
or when some randomness is present \cite{Sche03}.

The phenomenology just briefly described has been found to be valid in many bulk systems. It has only been realised
recently \cite{Plei04} that in semi-infinite critical systems similar dynamical scaling is also observed for surface
quantities. At the surface we can define, in complete analogy with a bulk system, surface autocorrelation and
autoresponse functions:
\BEQ 
C_1(t,s) = \left\langle \phi_1(t) \phi_1(s)\right\rangle \;\; , \;\;
R_1(t,s) = \left.\frac{\delta \langle\phi_1(t)\rangle}{\delta h_1(s)}\right|_{h_1=0} ~~(t > s)
\EEQ
where $\phi_1(t)$ is now the surface order parameter and $h_1$ is a field acting solely on the surface layer.
In the regime $t$, $s$, $t-s \gg t_{micro}$ we expect simple scaling forms:
\begin{eqnarray} 
C_1(t,s) & = & s^{-b_1} \, f_{C_1}(t/s) \label{C1} \\
R_1(t,s) & = & s^{-1-a_1} \, f_{R_1}(t/s) \label{R1}
\end{eqnarray}
and the scaling functions $f_{C_1}(y)$ and $f_{R_1}(y)$ should display a power-law behaviour
in the limit $y \longrightarrow \infty$:
\begin{equation} \label{fc1}
f_{C_1}(y) \sim y^{-\lambda_{C_1}/z} \;\; , \;\;
f_{R_1}(y) \sim y^{-\lambda_{R_1}/z}.
\end{equation}
Equations (\ref{C1}) - (\ref{fc1}) define the surface exponents $a_1$, $b_1$, $\lambda_{C_1}$ and $\lambda_{R_1}$.
In general these surface non-equilibrium exponents may take on values that differ from their bulk counterparts. This
is similar to what is known for static critical exponents where the surface values differ from the bulk ones \cite{Die97,Ple04a}.
Relations between the different non-equilibrium exponents can again be derived from general scaling arguments \cite{Plei04,CalGam05}.
Thus we obtain
\BEQ
a_1 = b_1 = (d-2 + \eta_\parallel)/z
\EEQ
where the static exponent $\eta_\parallel$ governs the decay of correlations parallel to the surface. For the surface autocorrelation
exponent $\lambda_{C_1} = \lambda_{R_1}$ one finds \cite{RiCz_95,Plei04b}
\BEQ
\lambda_{C_1} = \lambda_C + \eta_\parallel - \eta.
\EEQ
The scaling laws (\ref{C1}) - (\ref{fc1}) and the different relations between the non-equilibrium critical exponents have been 
verified in \cite{Plei04} for the critical semi-infinite Ising models in two and three dimensions. In \cite{CalGam05} the scaling 
forms in the presence of a surface have been discussed in more detail and calculations within the Gaussian
model have been presented.

One of the most intriguing aspects of surface critical phenomena is the presence of different surface universality
classes for a given bulk universality class \cite{Die97,Ple04a}. In this paper we study ageing phenomena in
semi-infinite spherical models with Dirichlet boundary conditions (corresponding to the so-called ordinary
transition where the bulk alone is critical) and with Neumann boundary conditions (corresponding to the
special transition point where both bulk and surface are critical). Besides investigating the surface out-of-equilibrium
dynamics at the critical point we also analyse the dynamical behaviour close to a surface in the ordered phase.
To our knowledge this is the first study of surface ageing phenomena in systems where phase ordering takes place.
As we shall see we thereby obtain for Dirichlet boundary conditions 
the unexpected result that the value of the non-equilibrium exponent $b_1$,
describing the dynamical scaling of the surface autocorrelation (\ref{C1}), is different from zero, the value
encountered in bulk systems undergoing phase ordering \cite{Bray94}.

The present work is on the one hand meant to close a gap in the study of kinetic spherical models \cite{Godreche2000},
which up to now have been restricted to systems with periodic boundary conditions. On the other hand our intention
is also to extend towards dynamics earlier investigations of the surface criticality of the spherical model 
\cite{Knops73,Bar73,Bar74,Sin75,Dan97a,Dan97b}. It has to be noted in this context that the static properties
of the critical semi-infinite spherical model with one spherical field have been shown to differ from those of the $O(N)$ model with 
$N \longrightarrow  \infty$ \cite{Bar74,Sin75}, even so both models are strictly equivalent in the bulk system \cite{Stan68,Kac71}.

The paper is organised in the following way. In Section 2 we present the kinetic model in a quite general
way, thus leaving open the possibility to consider different boundary  conditions in the different space
directions. In Sections 3 and 4 we discuss the dynamical scaling behaviour of surface correlation and response
functions, whereas in Section 5 we compute the surface fluctuation-dissipation ratio. Finally, Section 6
gives our conclusions.

\section{The model}
%
\subsection{General setting}
We consider a finite hypercubic system $\Lambda$ in $d$ dimensions
containing $\mathcal{N} = L_1
\times \cdots \times L_d$ sites where $L_{\nu}$ denotes the length
of the $\nu$-th edge \footnote{We set the lattice spacing equal
to one.}. To every lattice site $\vec{r}^T = (r_1,\ldots,r_d)$ we associate
a time-dependent classical spin variable $S(\vec{r},t)$. 
These spin variables are subjected to the mean spherical
constraint
\BEQ
\label{gl:constr}
\sum_{\vec{r} \in \Lambda} \langle S^2(\vec{r},t) \rangle =
\mathcal{N}
\EEQ
where the brackets indicate an average over the thermal noise.

As we are interested in the $d$-dimensional slap geometry we impose periodic boundary conditions
in all but one space direction. We then have in the $\nu$-th direction
(with $\nu = 2, \cdots, d$)
\BEQ
    \hspace{-1.5truecm}
    S( (r_1,\ldots,r_\nu + m \, L_\nu,\ldots,r_d),t ) = S(
    (r_1,\ldots,r_\nu,\ldots,r_d),t ) \qquad \mbox{for all} \; m
    \in \mathbb{Z}.
\EEQ
In the remaining direction we either consider Dirichlet boundary conditions,
corresponding to the ordinary transition, or Neumann boundary conditions,
corresponding to the special transition point. For Dirichlet boundary conditions 
we impose that
\BEQ
    S( (0,r_2,\ldots,r_d) ,t)= S(
    (L_1+1,r_2,\ldots,r_d),t ) = 0.
\EEQ
For Neumann boundary conditions, on the other hand, we have:
\begin{eqnarray}
    && S( (0,r_2,\ldots,r_d),t )  =  S(
    (1,r_2,\ldots,r_d),t ) \nonumber \\
    && S((L_1+1,r_2,\ldots,r_d),t )  = 
    S((L_1,r_2,\ldots,r_d),t ).
\end{eqnarray}
In the following quantities depending on the chosen boundary condition will be labeled by the superscript
$(p)$, $(d)$, and $(n)$ for periodic, Dirichlet, and Neumann boundary conditions, respectively.

The Hamiltonian can be written in the following general form
\BEQ \label{gl:hamil}
H = -\frac{1}{2}  J \vec{S}_{\Lambda}^{T} \cdot
\mathbf{Q}_{\Lambda}^{(\tau)}\cdot  \vec{S}_{\Lambda} +
\frac{1}{2} \lambda^{(\tau)} (t) \vec{S}_{\Lambda}^{T}\cdot \vec{S}_{\Lambda} -
\vec{S}_{\Lambda}^{T}\cdot \vec{h}_{\Lambda}
\EEQ
where the vector $\vec{S}_{\Lambda} := \left\{ S(\vec{r}) | \vec{r} \in \Lambda \right\}$
characterizes the state of the system. $J >0$ is the strength of the ferromagnetic
nearest neighbour couplings and is chosen to be equal to one in the following.
Furthermore, $\vec{h}_{\Lambda}:= \left\{ h(\vec{r}) | \vec{r} \in \Lambda \right\}$
where $h(\vec{r})$ is an external field acting on the spin at site
$\vec{r}$, whereas $\lambda (t)$
is the Lagrange multiplier ensuring the constraint (\ref{gl:constr}).
Finally, the interaction matrix $\mathbf{Q}_{\Lambda}^{(\tau)}$ (with
$\tau = (\tau_1, \cdots , \tau_d)$ characterizing the boundary conditions in the $d$ different
directions) is given by the tensor product \cite{Brankov}
\BEQ \label{gl:Q}
\mathbf{Q}_{\Lambda}^{(\tau)} = \bigotimes_{\nu=1}^d
(\Delta_{\nu}^{(\tau_{\nu})} + 2 \mathbf{E}_{\nu}). 
\EEQ
Here $\mathbf{E}_{\nu}$ is the unit matrix of dimension $L_{\nu}$,
whereas $\Delta_{\nu}^{(\tau_\nu)}$ is the discrete Laplacian in
the $\nu$-direction which depends on the boundary condition. 

It is convenient to parametrise the Langrange multiplier in the
following way \cite{Brankov}:
\[
\lambda^{(\tau)} (t) = \mu_{\Lambda}^{(\tau)}(\vec{\kappa}) + z^{(\tau)}(t)
\]
where $\mu_{\Lambda}^{(\tau)}(\vec{\kappa})$ is the largest eigenvalue 
of the interaction matrix $\mathbf{Q}_{\Lambda}^{(\tau)}$ (see Appendix A). In case of
periodic boundary conditions in {\it all} directions we recover the
usual expression
\[
\lambda^{(p)} (t) =  2 d + z^{(p)}(t).
\]

\subsection{Langevin equation}
In order to study the out-of-equilibrium dynamical behaviour of the semi-infinite
kinetic spherical model we prepare the system at time $t=0$ in a fully disordered infinite temperature
equilibrium state with vanishing magnetisation. We then bring the system in contact
with a thermal bath at a given temperature $T$ and monitor its temporal evolution. 
Assuming purely relaxational dynamics
(i.e. model A dynamics), the dynamics of the system is given by the following
stochastic Langevin equation:
\BEQ \label{gl:langevin_equ}
\frac{d}{dt}\vec{S}_\Lambda(t) = (\mathbf{Q}_{\Lambda}^{(\tau)} -
z^{(\tau)}(t)- \mu_{\Lambda}^{(\tau)}(\vec{\kappa})) \vec{S}_\Lambda(t) +
\vec{h}_\Lambda (t) + \vec{\eta}_\Lambda(t)
\EEQ
where $\vec{\eta}_\Lambda(t):= \left\{ \eta_{\vec{r}}(t) | \vec{r} \in \Lambda \right\}$ with
$\eta_{\vec{r}}(t)$ being a Gaussian white noise:
\[
\langle \eta_{\vec{r}}(t) \rangle  = 0 ~~~\mbox{and} ~~~
\langle \eta_{\vec{r}}(t) \eta_{\vec{r}'}(t') \rangle =  2 T \delta_{\vec{r},\vec{r}'} \delta(t-t').
\]
We further assume that the external field $\vec{h}_\Lambda$ is time-dependent.

Equation (\ref{gl:langevin_equ}) can easily be solved, yielding
\BEA \label{gl:sk}
\hspace{-2cm}
\vec{S}_{\Lambda}(t) &=& \exp \left(\int_0^t \D t'
(\mathbf{Q}_{\Lambda}^{(\tau)} - z^{(\tau)}(t') -
\mu_{\Lambda}^{(\tau)} (\vec{\kappa})) \right) \times
\Big[\vec{S}_{\Lambda}(0) \nonumber \\
\hspace{-2cm}&+&  \int_0^t \D t' \left(\exp \left( -\int_0^{t'} \D t''
(\mathbf{Q}_{\Lambda}^{(\tau)} - z^{(\tau)}(t'') -
\mu_{\Lambda}^{(\tau)}(\vec{\kappa}))\right) \right) (\vec{h}_{\Lambda}(t')
+ \vec{\eta}_{\Lambda}(t'))\Big].
\EEA
In order to proceed further we use a Fourier-transform-like
operation by defining
\BEQ \label{gl:fourier}
\wit{S}(\vec{k},t) := 
\sum_{\vec{r} \in \Lambda}\left( \prod_{\nu = 1}^d u^{(\tau_{\nu})}_{L_\nu}
(r_{\nu},k_{\nu})\right) S(\vec{r},t)
\EEQ
where the vectors $u^{(\tau_{\nu})}_{L_\nu}(r_{\nu},k_{\nu})$ are given in
Appendix A. We get the original vector back by the inverse
transformation
\BEQ
S(\vec{r},t) = \sum_{\vec{k} \in \Lambda} \left( \prod_{\nu = 1}^d {u_{L_\nu}^{(\tau_{\nu})}}^*(r_{\nu},k_{\nu}) \right)
\wit{S}(\vec{k},t).
\EEQ
The transformation (\ref{gl:fourier}) diagonalizes the symmetric matrix
$\mathbf{Q}_{\Lambda}^{(\tau)} - z(t) -
\mu_{\Lambda}^{(\tau)}(\vec{\kappa})$,
yielding
\BEA
\label{gl:mag_fourier}
\hspace{-1.5cm}
\wit{S}(\vec{k},t) = \frac{e^{-\omega^{(\tau)}(\vec{k}) t}}{\sqrt{g^{(\tau)}(t)}}
\left[ \wit{S}(\vec{k},0) + \int_0^t \D t' \left( \sqrt{g^{(\tau)}(t')}
e^{\omega^{(\tau)}(\vec{k})t'}(\wit{h}(\vec{k},t') +
\wit{\eta}(\vec{k},t'))\right)\right] 
\EEA
with
\BEQ
g^{(\tau)}(t) := \exp \left(2 \int_0^t \D u \,
z^{(\tau)}(u)\right)
\EEQ
and
\BEQ \label{gl:omega}
\omega^{(\tau)}(\vec{k}) := -\mu_{\Lambda}^{(\tau)}(\vec{k}) +
\mu_{\Lambda}^{(\tau)}(\vec{\kappa}).
\EEQ

We now take the limit of the semi-infinite system which extends from $- \infty$ to $\infty$
in the directions parallel to the surface, whereas in the remaining direction the coordinate $r_1$
takes on only positive values. In order to stress the existence of the special direction we set
$\vec{r}^T = (r, \vec{x}^T)$ with $r:= r_1$ and $\vec{x}^T:=(r_2, \ldots, r_d)$. The corresponding vector
in reciprocal space is then written as $\vec{k}^T = (k, \vec{q}^T)$ with $k:= k_1$ and
$\vec{q}^T:=(k_2, \ldots, k_d)$. As a consequence of the semi-infinite volume limit sums have to
be replaced by integrals in the following way:
\begin{itemize}
\item 
$\frac{2}{L_1}\sum\limits_{k_1=1}^{L_1} (\ldots) \longrightarrow
\frac{2}{\pi} \int\limits_0^\pi \D k (\ldots)$ in the direction perpendicular to the surface,
\item 
$\frac{1}{L_{\nu}} \sum\limits_{k_{\nu}=1}^{L_{\nu}} (\ldots)
\longrightarrow \frac{1}{2 \pi} \int\limits_{-\pi}^{\pi} \D k
(\ldots)$ in all other directions,
\end{itemize}
where the integration limits follow from 
Appendix A.
It has to be noted that in the semi-infinite volume limit
the largest eigenvalue of the interaction matrix (\ref{gl:Q}) is given by
$ \mu_{\Lambda}^{(\tau)}(\vec{\kappa}) = 2 d$
irrespective of the chosen boundary condition. We therefore rewrite
(\ref{gl:omega}) in the following way:
\BEQ
\omega^{(\tau)}(\vec{k}) = \omega(k,\vec{q}) = \wit{\omega}(k)+\wit{\wit{\omega}}(\vec{q})
\EEQ
with
\BEQ
\wit{\omega}(k) = 2 \left( 1 -  \cos k \right) ~~~ \mbox{and} ~~~ \wit{\wit{\omega}}(\vec{q}) = 2 \sum\limits_{\nu=2}^d \left(
1 - \cos k_\nu \right).
\EEQ
where we used the explicit expressions for $\mu^{(\tau )}_{\Lambda}(\vec{k})$ (see Appendix A).

As usual we prepare the system at time $t=0$ in a completely uncorrelated initial state with 
vanishing magnetization. We then have at $t=0$ in reciprocal space
\BEQ
\langle \wit{S}(\vec{k},0) \wit{S}(\vec{k}',0) \rangle
    =  (2 \pi)^{d-1} \frac{\pi}{2} \delta^{d-1}(\vec{q}+\vec{q}') C^{(\tau)}(k,k') 
\EEQ
where the quantity $C(k,k')$ is given by
\BEA
 C^{(\tau)}(k,k')  = && 
    \left\{ \begin{array}{ll}
    \sum_{r=1}^\infty \sin ( r k ) \sin (r k') & \tau = d \\
    \sum_{r=1}^\infty \cos ( (r-\frac{1}{2}) k ) \cos ( (r-\frac{1}{2}) k' ) & \tau = n
    \end{array}
    \right.
\EEA    
When we transform this expression back into direct space it just gives a decorrelated intial correlator, 
as is easily checked.

Before deriving the scaling functions of dynamical two-point functions we have to pause briefly
in order to discuss possible implementations of the mean spherical constraint in the
semi-infinite geometry. Our starting point is Eq.\ (\ref{gl:constr}) which should hold true 
at all times. However, in the semi-infinite geometry translation invariance in the direction
perpendicular to the surface is broken. One can therefore not assume {\it a priori} that the
spins in different layers should be treated on an equal footing. In the past investigations of
the static properties of the semi-infinite spherical model either considered one global spherical 
field \cite{Bar74,Dan97a} or introduced besides the bulk spherical field an additional spherical
field for the surface layer \cite{Sin75,Dan97b}. Both cases belong to universality classes which
differ from that of the $O(N)$ model with $N \longrightarrow \infty$.
In studies of {\it static}
quantities it has been proposed that the spherical constraint could be realised in the semi-infinite geometry
by imposing in the Hamiltonian a different spherical field for every layer \cite{Knops73}, thus 
yielding the universality class of the semi-infinite $O(N)$ model with $N \longrightarrow \infty$.
For dynamical
quantities, however, this prescription leads to the unwanted effect that the matrices 
appearing in the exponential in Eq.\ (\ref{gl:sk}) do not commute anymore, thus making an analytical 
treatment of the dynamical properties of the semi-infinite model prohibitively difficult. We
therefore made the choice to retain only one spherical field, similar to what is done in the
bulk system. This then yields the Equations (\ref{gl:mag_fourier}) - (\ref{gl:omega}) that are at
the centre of our considerations. It has to be stressed that this implementation of
the mean spherical constraint is an integral part of the model studied in this paper.

\section{The surface autocorrelation function}
The two-time spin-spin correlation function is defined by
\BEQ
C^{(\tau)}(\vec{r},\vec{r}\,'; t,s) = \langle S(\vec{r},t) S(\vec{r}',s) \rangle
\EEQ
with $\vec{r}^T = (r, \vec{x}^T)$ and $\vec{r}'\;^T = (r, \vec{x}'\;^T)$. Here $(\tau)$ again
characterises the chosen boundary condition. In the semi-infinite system correlation functions
are expected to behave differently close to the surface and inside the bulk. Of special
interest is the surface autocorrelation
\BEQ \label{gl:autocorr}
C_1^{(\tau)}(t,s) = C^{(\tau)}((1,\vec{x}),(1,\vec{x});t,s)
\EEQ
where we follow the convention to attach a subscript 1 to surface related quantities.

{}From Eq.\ (\ref{gl:mag_fourier}) we directly obtain the following expression for
the correlation function in reciprocal space
\BEA
C^{(\tau)}(\vec{k},\vec{k}';t,s) 
&=& \frac{e^{-\omega(k,\vec{q}) t - \omega(k',\vec{q}')
s}}{\sqrt{g^{(\tau)}(t) g^{(\tau)}(s)}}(2 \pi)^{d-1} \frac{\pi}{2} \delta^{d-1} (\vec{q}+\vec{q}')
\Big[ C^{(\tau)}(k,k') \nonumber \\ 
&+&  2 T  \delta(k-k') \int_0^t \D t' e^{2
\omega(k,\vec{q}) t'} g^{(\tau)}(t') \Big]
\EEA
where we have set the external field to zero. Transforming back to real space we 
obtain
\BEA \label{Creal}
C^{(\tau)}((r,\vec{x}),(r',\vec{x});t,s) &=&  \frac{1}{\sqrt{g^{(\tau)}(t) g^{(\tau)}(s)}}
\left[ f^{(\tau)}\left(r,r';\frac{t+s}{2}\right) \right. \nonumber \\
&+&\left. 2 T \int_0^s \D u \, f^{(\tau)}\left(r,r';\frac{t+s}{2} -u \right) g^{(\tau)}(u) \right]
\EEA
where a boundary dependent function $f^{(\tau)}$ has been defined, with
\BEA
f^{(d)}(r,r';t) &=& \frac{2}{\pi}  \int \frac{\D^{d-1} \vec{q}}{(2 \pi)^{(d-1)}} \,
e^{-2 \wit{\wit{\omega}}(\vec{q})t} \\ & &\int_0^\pi \D k
\sin(r\cdot k) \sin (r' \cdot k) e^{-2 \wit{\omega}(k) t}
\nonumber \\
& = & (e^{-4 t} I_0(4 t))^{d-1}\cdot e^{-4 t}
(I_{r-r'}(4 t) - I_{r+r'}(4 t))
\EEA
and
\BEA
f^{(n)}(r,r';t) &=& \frac{2}{\pi}  \int \frac{\D^{d-1}\vec{q}}{ (2 \pi)^{(d-1)}}
e^{-2 \wit{\wit{\omega}}(\vec{q})t} \nonumber \\ & &\int_0^\pi \D k
\cos( (r-1/2)\cdot k) \cos ( (r'-1/2) \cdot k) e^{-2 \wit{\omega}(k) t}
\nonumber \\
& = & (e^{-4 t} I_0(4 t))^{d-1}\cdot e^{-4 t}
(I_{r-r'}(4 t) + I_{r+r'-1}(4 t)) 
\EEA
with the modified Bessel functions $I_\nu(u)$.
In the long time limit $t \longrightarrow \infty$ we get
\BEA
f^{(d)}(r,r';t) &\stackrel{t \rightarrow \infty}{\approx}& 4
\pi (8 \pi t)^{-\frac{d+2}{2}} r \cdot r' \label{gl:fd} \\
f^{(n)}(r,r';t) &\stackrel{t \rightarrow \infty}{\approx}& 2
(8 \pi t)^{-\frac{d}{2}}  \label{gl:fn}
\EEA
where the well-known asymptotic expansion
\BEQ
\label{gl:bessel_asymptotic}
I_{\nu}(u)  \stackrel{u \rightarrow \infty}{\approx}
\frac{e^u}{\sqrt{2 \pi u}} \left( 1 - \frac{\nu^2 - 1/4}{2
u} + O(1/u^2)\right) 
\EEQ
of the Bessel function has been used.

The function $g^{(\tau)}(t)$ is obtained as the solution of a nonlinear Volterra
equation
\BEQ
\label{gl:constraint2}
g^{(\tau)}(t) = \lim_{L \rightarrow \infty} \frac{1}{L}
\sum_{r=1}^L \left( f^{(\tau)}(r,r;t) + 2 T \int_0^t \D u
f^{(\tau)}(r,r;t-u) g^{(\tau)}(u) \right)
\EEQ
that follows directly from the implementation of the mean spherical constraint
through the requirement
\BEQ
\label{gl:requirement}
\lim_{L \rightarrow \infty} \frac{1}{L} \sum_{r=1}^L
C^{(\tau)}((r,\vec{0}),(r,\vec{0});t,t) = 1 
\EEQ
for all times $t$. Inserting the expressions (\ref{gl:fd}) and (\ref{gl:fn}) for $f^{(\tau)}$
into Eq. (\ref{gl:constraint2}) we observe that we end up with the known bulk equation \cite{Cug1995}
\BEQ
g(t) = \hat{f}(t) + 2 T \int_0^t \D u \hat{f}(t-u)g(u)
\EEQ
with
\BEQ
\label{gl:const_function}
\hat{f}(t) := \int_{-\pi}^{\pi} \frac{\D^{d-1} \vec{q}}{(2 \pi)^{d-1}}
\int_{-\pi}^{\pi} \frac{\D k}{2 \pi} e^{-2 \omega(k,\vec{q}) t} = (e^{-4 t} I_0(4 t))^d,
\EEQ
indepedently of whether we choose Dirichlet or Neumann boundary conditions. We can therefore
drop in the following the superscript $(\tau)$ when dealing with the function $g$. 
The asymptotic behavior of $g$ has been discussed in detail in \cite{Godreche2000}.
We recall these results in Appendix B as they are needed in the following.

Everything is now in place for a discussion of the behavior of the two-time correlations in the
dynamical scaling regime $t$, $s$, $t-s \gg 1$. We shall in the following focus on the surface autocorrelation
function. Other correlations (for example the correlation between surface and bulk spins)
can be discussed along the same line.
The surface autocorrelation function is readily obtained from Eq.\ (\ref{Creal}) by setting $r = r' = 1$. 
\begin{itemize}
\item \underline{$T < T_C$:} From renormalisation arguments we know that 
temperature is an irrelevant variable below the critical point and can be set to zero, and only the
first term in (\ref{Creal}) contributes in the scaling limit \cite{Godreche2000}. For both boundary conditions we obtain a 
dynamical scaling behavior, as
\BEA
\label{gl:res:ucrit_corr1}
C^{(d)}(t,s) & = & 2^{\frac{d}{2}} M^2_{eq}
\, s^{-1} \left(\frac{t}{s} \right)^{\frac{d}{4}} \left(
\frac{t}{s} + 1 \right)^{-(\frac{d}{2}+1)}, \\
\label{gl:res:ucrit_corr0}
C^{(n)}(t,s) & = & 2^{\frac{d}{2}+1} M^2_{eq}  \left( \frac{t}{s} \right)^{\frac{d}{4}}
\left( \frac{t}{s}  + 1 \right)^{-\frac{d}{2}}.
\EEA
where $M_{eq}^2 = 1 - T/T_C$ \cite{Baxter82}.
The non-equilibrium exponents can then be read off directly (recall $z=2$):
\BEA \label{gl:exp_tltc}
b_1^{(d)} = 1, \qquad b_1^{(n)} = 0
, \qquad \lambda_{C_1}^{(d)} = \frac{d}{2}
+ 2,\qquad \lambda_{C_1}^{(n)} = \frac{d}{2}.
\EEA

A few comments are now in order. When comparing the values (\ref{gl:exp_tltc}) with those obtained for the
bulk system \cite{Janssen89,Godreche2000}, see Table \ref{table1}, we observe that for Neumann boundary conditions
the non-equilibrium exponents take on exactly the same values than in the bulk.
This is not only observed for quenches
to temperatures below the critical point but also for quenches to the critical point, see below, and also 
holds for the exponents related to the response function, as discussed in the next Section. Interestingly,
the agreement between Neumann and periodic boundary conditions (i.e. bulk systems) also extends to the scaling
functions of two-time quantities which are identical, up to an unimportant numerical factor.
\footnote{This behavior may be compared to that of the critical semi-infinite Ising model at the special transition point
where the values of the non-equilibrium surface and bulk exponents and the surface and bulk scaling
functions are found to disagree \cite{Plei04}. Similarly, the surface autocorrelation exponent $\lambda_{C_1}$
for the $O(N)$
model in the limit $N \longrightarrow \infty$ differs at the special transition point from the bulk
autocorrelation exponent $\lambda_C$ \cite{Maj96}. Differences at the special transition point between the semi-infinite spherical and 
the Ising and $O(N)$ ($N \longrightarrow \infty$) models already show up 
in the static properties as the surface and bulk order parameter
exponents are identical for the spherical model but 
differ for the other two models \cite{Ple04a}.}

\begin{table}[h]
\begin{center}
\begin{tabular}{|c|c|c|c|}  \hline
 & periodic ($\tau$ = p) & Dirichlet ($\tau$ = d) & Neumann ($\tau$ = n) \\ \hline
\multicolumn{4}{|c|}{$T<T_c$, $d > 2$} \\ \hline
$b_1^{(\tau)}$ & $0$ & $1$ & $0$\\ \hline
$a_1^{(\tau)}$ & $\frac{d}{2}-1$ & $\frac{d}{2}$ & $\frac{d}{2}-1$ \\ \hline
$\lambda_{C_1}^{(\tau)} = \lambda_{R_1}^{(\tau)}$ & $\frac{d}{2}$ & $\frac{d}{2}+2$ & $\frac{d}{2}$ \\ \hline
$X_\infty^{(\tau)}$ & $0$ & $0$ & $0$ \\ \hline
\multicolumn{4}{|c|}{$T=T_c$, $2 < d < 4$} \\ \hline
$b_1^{(\tau)}$ & $\frac{d}{2}-1$ & $\frac{d}{2}$ & $\frac{d}{2}-1$ \\ \hline
$a_1^{(\tau)}$ & $\frac{d}{2}-1$ & $\frac{d}{2}$ & $\frac{d}{2}-1$ \\ \hline
$\lambda_{C_1}^{(\tau)}= \lambda_{R_1}^{(\tau)}$ & $\frac{3d}{2}-2$ & $\frac{3d}{2}$ & $\frac{3d}{2}-2$ \\ \hline
$X_\infty^{(\tau)}$ & $1 - \frac{d}{2}$ & $1 - \frac{d}{2}$ & $1 - \frac{d}{2}$ \\ \hline
\multicolumn{4}{|c|}{$T=T_c$, $d > 4$} \\ \hline
$b_1^{(\tau)}$ & $\frac{d}{2}-1$ & $\frac{d}{2}$ & $\frac{d}{2}-1$ \\ \hline
$a_1^{(\tau)}$ & $\frac{d}{2}-1$ & $\frac{d}{2}$ & $\frac{d}{2}-1$ \\ \hline
$\lambda_{C_1}^{(\tau)}=\lambda_{R_1}^{(\tau)}$ & $d$ & $d+2$ & $d$ \\ \hline
$X_\infty^{(\tau)}$ & $\frac{1}{2}$ & $\frac{1}{2}$ & $\frac{1}{2}$ \\ \hline
\end{tabular}
\end{center}
\caption{Ageing exponents and fluctuation-dissipation ratio for periodic (bulk system), Dirichlet and Neumann
boundary conditions. \label{table1}}
\end{table}

The situation is different when considering open (i.e.\ Dirichlet) boundary conditions. The values of the
non-equilibrium exponents are distinct from the values of their bulk counterparts and the scaling functions are
different, too. The larger value of the surface autocorrelation exponent $\lambda_{C_1}^{(d)}$ thereby reflects
the increased disorder at the surface due to the absence of neighbouring spins. Remarkably, the exponent
$b_1^{(d)}$ is found to be different from zero, the value usually encountered when studying phase ordering in 
a bulk system. It is an open question whether this surprising observation is unique to the special situation
of the spherical model or whether it is encountered in other systems as for example the semi-infinite Ising
model quenched to temperatures below $T_c$.

The two types of dynamical scaling behaviour encountered in the semi-infinite spherical model quenched below
the critical point are illustrated in Figure \ref{fig1} in three dimensions. The analytical curves are given by
the expressions (\ref{gl:res:ucrit_corr1}) and (\ref{gl:res:ucrit_corr0}) whereas the symbols,
corresponding to different waiting times, are obtained from
the direct numerical evaluation of Equation (\ref{Creal}). It is obvious that we are well within the dynamical 
scaling regime even for the smallest waiting time considered. Furthermore this confirms {\it a posteriori}
that it was justified to drop the second term in $(\ref{Creal})$.

\begin{figure}[ht]
  \vspace{0.5cm}
  \centerline{\epsfxsize=5.5in\epsfclipon\epsfbox
  {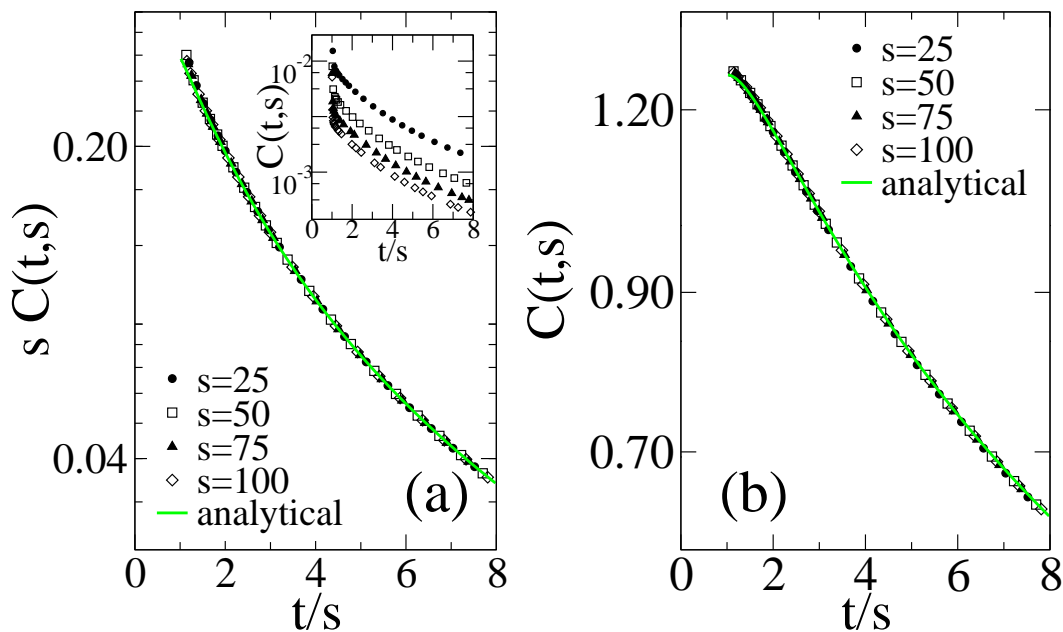}
  }\caption{Scaling plots of the autocorrelation function
  for the case $T < T_C$ in three dimensions: 
  (a) for Dirichlet boundary conditions with $b_1^{(d)}=1$,
  (b) for Neumann boundary conditions with $b_1^{(n)}=0$.
  The inset in (a) shows that no data collapse is observed for the
  unscaled data.}
  \label{fig1}
\end{figure} 
\item \underline{$T = T_C$ and $2 < d < 4$:}
The behavior of $C_1^{(\tau)}(t,s)$ in the regime $t$, $s$, $t-s \gg 1$ follows from the 
insertion of the asymptotic expressions for $g$  and for $f^{(\tau)}$ into Eq.\ (\ref{Creal}).
One may wonder whether the use of the asymptotics for $g$ in the integrand of Eq.\ (\ref{Creal})
does not cause problems at the lower integration bound. As has been shown in \cite{Zippold2000}
there exists for the spherical model a time scale $t_P \sim s^\zeta$ with $ 0 < \zeta < 1$ such that for
times larger than $t_P$ one is well within the ageing regime. Replacing $g$ by its asymptotic value
$g_{age}$ for $u > t_P$, the integral of Eq.\ (\ref{Creal}) can be written in the following way:
\BEA
& & \hspace{-2.0cm} \int_0^s \D u \, f^{(\tau)}(1,1;(t+s)/2 - u) g(u) \nonumber \\
& & \hspace*{-0.5cm} =  \int_0^{t_P} \D u g(u) f^{(\tau)}(1,1;(t+s)/2 - u)  + 
 \int_{t_P}^s \D u \, g_{age}(u) f^{(\tau)}(1,1;(t+s)/2 -u)\nonumber \\
 & = & W^{(\tau)}(t,s,t_P)  +  s \int_0^1 dv \, g_{age}(sv) f^{(\tau)}(1,1;(t+s)/2 - sv) \label{W}
\EEA
where in the last line we have assumed $s$ to be large. An upper bound for the first term
$W^{(\tau)}(t,s,t_P) = \int_0^{t_P} \D u g(u) f^{(\tau)}(1,1;(t+s)/2 - u)$ is given by
\BEA
& &|W_P^{(\tau)}(t,s,t_P)| \leq t_P \max_{\tau \in [0,t_P]} | g(\tau)
f^{(\tau)} (1,1;(t+s)/2 - \tau) |\nonumber  \\
& \stackrel{s \gg 1}{\approx}& t_P \cdot  \max_{\tau \in
[0,t_P]} | g(\tau) |s^{-\frac{d}{2}} \left(4 \pi \left(
\frac{1}{2} (\frac{t}{s} + 1) - \frac{t_P}{s} \right)
\right)^{-\xi^{(\tau)}}
\EEA
where $\xi^{(d)} = \frac{d}{2}+1$ and $\xi^{(n)} =
\frac{d}{2}$. As $t_P \sim s^\zeta$, this upper bound disappears in the scaling limit faster than the second
contribution in Eq.\ (\ref{W}). We therefore drop $W^{(\tau)}(t,s,t_P)$ in the following and end up with 
the expression
\BEA
\label{gl:autocorrelator2}
C_1^{(\tau)}(t,s) &=&
\frac{1}{\sqrt{g_{age}(t)g_{age}(s)}} \left[ f^{(\tau)}
\left(1,1;\frac{t+s}{2}\right)\right. \nonumber \\
& + & \left. 2 T \int_0^s \D u
f^{(\tau)}\left(1,1;\frac{t+s}{2}-u\right)g_{age}(u) \D u \right]
\EEA
in the asymptotic regime. It turns out that the thermal
part of the expression (\ref{gl:autocorrelator2}) is the
leading one at the critical points in any dimensions, so that
the first term can be disregarded. For $2 < d < 4$ we then obtain
again dynamical scaling behavior, as shown in Figure \ref{fig2}, as:
\BEA
\label{gl:res:crit_corr1}
C^{(d)}(t,s) & = &  \frac{4 T_C (4
\pi)^{-(\frac{d}{2})}}{d-2} s^{-\frac{d}{2}} \left[
\left( \frac{t}{s}\right)^{1-\frac{d}{4}}
\left(\frac{t}{s}+1\right)^{-1}  \left( \frac{t}{s}- 1
\right)^{-\frac{d}{2}} - \right. \nonumber \\
& & \left. \frac{4}{d} \left( \frac{t}{s}
\right)^{1-\frac{d}{4}}
\left(\frac{t}{s}+1 \right)^{-2} \left(\frac{t}{s} -1
\right)^{-\frac{d}{2}} \right], \\
\label{gl:res:crit_corr0}
C^{(n)}(t,s) & = & \frac{8 (4
\pi)^{-\frac{d}{2}}\cdot T_C }{(d-2)} s^{-(\frac{d}{2}-1)}\left(\frac{t}{s}\right)^{1-\frac{d}{4}}
\left(\frac{t}{s}-1\right)^{1-\frac{d}{2}} \left(\frac{t}{s} + 1 \right)^{-1}
\EEA
yielding the non-equilibrium critical exponents
\BEA
b_1^{(d)} = \frac{d}{2}, \qquad b_1^{(n)} =
\frac{d}{2}-1, \qquad \lambda_{C_1}^{(d)} = \frac{3}{2}d,
\qquad \lambda_{C_1}^{(n)} = \frac{3}{2}d - 2.
\EEA
The values $b_1^{(\tau)}$ agree with the expression
\BEQ \label{gl:b1}
b_1^{(\tau)} = b + \frac{2 (\beta_1^{(\tau)} - \beta)}{\nu z}
\EEQ
expected from general scaling considerations \cite{CalGam05}. Eq.\ (\ref{gl:b1}) is readily verified in three dimensions
where $b= \frac{2 \beta}{\nu z} = \frac{1}{2}$, whereas $\beta$ and $\beta_1^{(\tau)}$,
which are the usual static equilibrium
critical exponents describing the vanishing of the bulk and surface magnetisations on approach to the critical point,
take on the values $\beta = \frac{1}{2}$ and $\beta_1^{(d)}= \frac{3}{2}$, $\beta_1^{(n)}= \frac{1}{2}$.
Finally, the static exponent $\nu =1$ in three dimensions.

As a final remark, let us note that the tendency of the surface correlations to decay faster in time 
than the bulk correlations
is mirrored at the ordinary transition by the larger value of the non-equilibrium autocorrelation exponent $\lambda_{C_1}^{(d)}$.

\begin{figure}[ht] 
  \vspace{0.5cm}
  \centerline{\epsfxsize=5.5in\epsfclipon\epsfbox
  {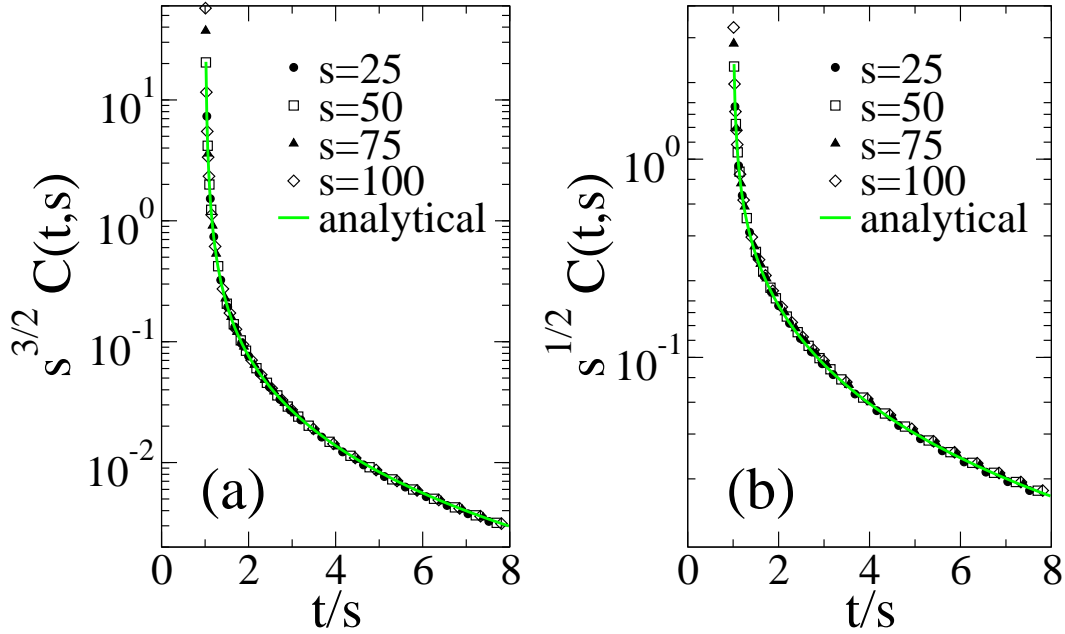}
  }\caption{Scaling plots of the autocorrelation function for
  the case $T=T_C$ in three dimensions: (a) at the ordinary
  transition (Dirichlet boundary conditions), (b) at the special transition point (Neumann boundary conditions).} 
  \label{fig2}
\end{figure} 

\item \underline{$T = T_C$ and  $d > 4$:}
In this case we obtain
\BEA
\label{gl:res:critmf_corr1}
\hspace{-1.0cm}
C^{(d)}(t,s) & = & 
\frac{2 (4 \pi)^{-\frac{d}{2}}}{d} \cdot
T_C \cdot s^{-\frac{d}{2}} \left(
\left(\frac{t}{s}-1\right)^{-\frac{d}{2}} -
\left(\frac{t}{s}+1\right)^{-\frac{d}{2}} \right) \\
\label{gl:res:critmf_corr0}
\hspace{-1.0cm}
C^{(n)}(t,s) & = & \frac{4 (4 \pi)^{-\frac{d}{2}}}{d-2}
\cdot T_C \cdot s^{-(\frac{d}{2}-1)} \left(
\left(\frac{t}{s}-1\right)^{1-\frac{d}{2}} -
\left(\frac{t}{s}+1\right)^{1-\frac{d}{2}} \right).
\EEA
Again dynamical scaling is found where the non-equilibrium critical exponents now take on the values
\BEA
b_1^{(d)} = \frac{d}{2}, \qquad b_1^{(n)} =
\frac{d}{2}-1, \qquad \lambda_{C_1}^{(d)} = d + 2,
\qquad \lambda_{C_1}^{(n)} = d.
\EEA
Also here (\ref{gl:b1}) is verified with the values $b = \frac{d}{2}-1$, $\beta_1^{(d)} = 1$, $ \beta_1^{(n)} = \frac{1}{2}$, $\nu = \frac{1}{2}$ and $\beta = \frac{1}{2}$.
\item \underline{$T > T_C$:} 
We then have
\BEA
C_1^{(\tau)}(t,s) & = & e^{-(t+s)/(2 \tau_{eq})} \left[
f^{(\tau)}\left(1,1;\frac{t+s}{2} \right) \right.\nonumber \\ &+& \left.2 T \int_0^s \D u
f^{(\tau)} \left(1,1;\frac{t+s}{2}-u \right) g_{age}(u) \right].
\EEA
This amounts to an exponential decrease and the leading
expression rapidly develops a dependence only on $t-s$ for 
$s \rightarrow \infty$ and $t-s$ fixed \cite{Godreche2000} .
\end{itemize}
Inspection of Table \ref{table1} reveals the oddity that the values of $a_1$ and $b_1$ for Dirichlet
and Neumann boundary conditions always exactly differ by one. This may again be compared with the critical semi-infinite
Ising model. We there have $a_1 = b_1 = 2 \beta_1/(\nu z)$ \cite{Plei04,CalGam05} with the values $\beta_1 = 0.80$
at the ordinary transition and $\beta_1 \approx 0.23$ at the special transition point \cite{Ple04a}, yielding different
values for $a_1$ and $b_1$ in that case, too. However, the exact value one for the difference seems to be a property
of the spherical model.

\section{The response function}
The response function is defined as usual:
\BEQ
R(\vec{r},\vec{r}';t,s) := \left. \frac{\delta \langle
S(\vec{r},t) \rangle }{\delta h(\vec{r}',s)
} \right|_{h=0}
\EEQ
where $h(\vec{r}',s)$ is a magnetic field acting at time $s$ on the spin located at lattice site $\vec{r}'$.
Starting from expression (\ref{gl:mag_fourier}) and assuming spatial
translation invariance parallel to the surface, we find
\BEQ
R((k,\vec{q}),(k',\vec{q});t,s) = \sqrt{\frac{g(s)}{g(t)}}
e^{-\omega(k,\vec{q})(t-s)} \delta(k-k').
\EEQ
The chosen boundary condition enters when transforming back to real space, yielding
\BEA
\label{gl:response_res1}
R^{(d)}((r,\vec{x}),(r',\vec{y});t,s) &=& \sqrt{\frac{g(s)}{g(t)}}
e^{-2(t-s)d} \left(I_{\vec{x}-\vec{y}}(2(t-s))\right)^{d-1}
\nonumber \\ &\times& \left( I_{r'-r}(2(t-s)) -
I_{r'+r}(2(t-s))\right) \\
\label{gl:response_res0}
R^{(n)}((r,\vec{x}),(r',\vec{y});t,s) &=& \sqrt{\frac{g(s)}{g(t)}}
e^{-2(t-s)d} \left(I_{\vec{x}-\vec{y}}(2(t-s))\right)^{d-1}
\nonumber \\ &\times& \left( I_{r'-r}(2(t-s)) +
I_{r'+r-1}(2(t-s))\right).
\EEA
We refrain from giving a full discussion of the response function, but focus instead on the
surface autoresponse function $R_1^{(\tau)}(t,s) = R^{(\tau)}((1,\vec{x}),(1,\vec{x});t,s)$ that
describes the answer of the surface at a certain position to a perturbation at the same site at an earlier time.
With the large $t$ behaviour (\ref{gl:bessel_asymptotic}) of the Bessel function we then find
\BEA
\label{gl:response1}
R_1^{(d)}(t,s) & = & 4 \pi
\sqrt{\frac{g(s)}{g(t)}} (4 \pi (t-s))^{-(\frac{d}{2}+1)} \\
\label{gl:response2}
R_1^{(n)}(t,s) & = & 2 \sqrt{\frac{g(s)}{g(t)}} (4
\pi (t-s))^{-\frac{d}{2}}.
\EEA
It is evident from these expressions that the ageing behaviour of the surface autoresponse is again determined
by the asymptotics $g_{age}$ of the function $g$ given in the Appendix B. 
As for the autocorrelation function we have to distinguish four different cases:
\begin{itemize}
  
   \item \underline{$T < T_C$:} Here we have
       \BEA
       \label{gl:res:ucrit_resp1}
           R_1^{(d)}(t,s) & = & (4 \pi)^{-\frac{d}{2}}
	       \left( \frac{t}{s} \right)^{\frac{d}{4}}
	           s^{-(\frac{d}{2}+1)} \left( \frac{t}{s}-1
		       \right)^{-(\frac{d}{2}+1)} \\
       \label{gl:res:ucrit_resp0}
           R_1^{(n)}(t,s) & = & 2 (4 \pi)^{-(\frac{d}{2})}
	       \left( \frac{t}{s} \right)^{\frac{d}{4}}
	           s^{-\frac{d}{2}} \left( \frac{t}{s}-1
		       \right)^{-\frac{d}{2}}.
       \EEA
       For both choices of boundary conditions we observe dynamical scaling (see Figure \ref{fig3}) with the exponents
       \BEA
           a_1^{(d)} = \frac{d}{2}, \qquad a_1^{(n)} =
	       \frac{d}{2}-1, \qquad \lambda_{R_1}^{(d)} = \frac{d}{2}
	           + 2,\qquad \lambda_{R_1}^{(n)} = \frac{d}{2}.
		       \EEA
\begin{figure}[h]
  \vspace{0.5cm}
  \centerline{\epsfxsize=5.5in\epsfclipon\epsfbox
  {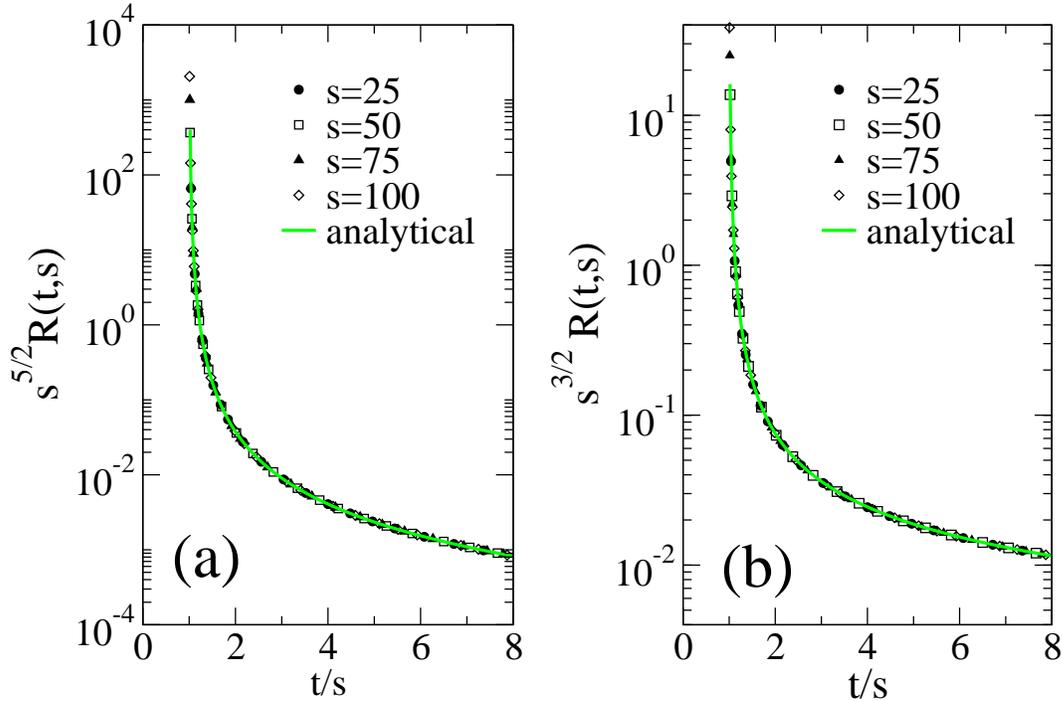}
  }\caption{Scaling plots of the autoresponse function
  for the case $T < T_C$ in three dimensions: 
  (a) for Dirichlet boundary conditions with $a_1^{(d)}=\frac{3}{2}$,
    (b) for Neumann boundary conditions with $d_1^{(n)}=\frac{1}{2}$.}
    \label{fig3}
\end{figure} 
   \item \underline{$T = T_C$ and $2 < d < 4$:} We find again dynamical scaling, as
   \BEA
   \label{gl:res:crit_resp1}
   R_1^{(d)}(t,s) & = & (4 \pi)^{-\frac{d}{2}}
       \left( \frac{t}{s} \right)^{-\frac{d}{4}+1}
           s^{-(\frac{d}{2}+1)} \left( \frac{t}{s}-1
	       \right)^{-(\frac{d}{2}+1)} \\
   \label{gl:res:crit_resp0}
       R_1^{(n)}(t,s) & = & 2 (4 \pi)^{-\frac{d}{2}}
           \left( \frac{t}{s} \right)^{-\frac{d}{4}+1}
	       s^{-\frac{d}{2}} \left( \frac{t}{s}-1
	           \right)^{-\frac{d}{2}}
  \EEA
with the following nonequilibrium critical exponents:
\BEA
    a_1^{(d)} = \frac{d}{2}, \qquad a_1^{(n)} =
        \frac{d}{2}-1, \qquad \lambda_{R_1}^{(d)} = \frac{3}{2}d,
	    \qquad \lambda_{R_1}^{(n)} = \frac{3}{2}d - 2.
	        \EEA
  \item \underline{$T = T_C$ and $d > 4$:} This case yields the expressions
      \BEA
      \label{gl:res:critmf_resp1}
          R_1^{(d)}(t,s) & = & (4 \pi)^{-\frac{d}{2}}
	      s^{-(\frac{d}{2}+1)} \left( \frac{t}{s}-1
	          \right)^{-(\frac{d}{2}+1)} \\
      \label{gl:res:critmf_resp0}
          R_1^{(n)}(t,s) & = & 2 (4 \pi)^{-\frac{d}{2}}
	      s^{-\frac{d}{2}} \left( \frac{t}{s}-1
	          \right)^{-\frac{d}{2}}.
      \EEA
      Again dynamical scaling is found, the exponents now taking on the values
      \BEA
          a_1^{(d)} = \frac{d}{2}, \qquad a_1^{(n)} =
	      \frac{d}{2}-1, \qquad \lambda_{R_1}^{(d)} = d + 2,
	          \qquad \lambda_{R_1}^{(n)} = d.
		      \EEA
This dynamical scaling behaviour is illustrated in Figure \ref{fig4} in five dimensions.
\begin{figure}[t]
  \vspace{0.5cm}
  \centerline{\epsfxsize=5.5in\epsfclipon\epsfbox
  {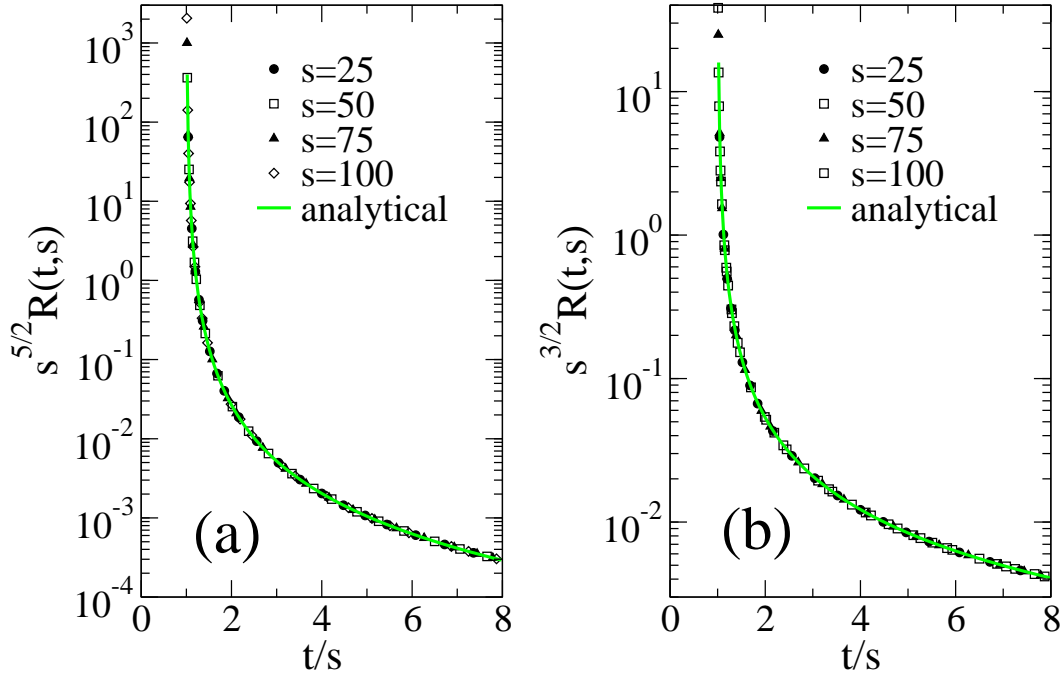}
  }\caption{Scaling plots of the autoresponse function for
  the case $T=T_C$ in five dimensions: a) at the ordinary
    transition (Dirichlet boundary conditions), (b) at the special transition point (Neumann boundary conditions).}
    \label{fig4}
\end{figure} 
\item \underline{$T > T_C$:} Due to the exponential
      behaviour of $g_{age}(t)$, $R(t,s)$ disappears exponentially:
      \BEA
      R_1^{(d)}(t,s) & = & 4 \pi e^{-(t-s)/\tau_{eq}} (4 \pi
          (t-s))^{-(\frac{d}{2}+1)} \\
      R_1^{(n)}(t,s) & = & 2 e^{-(t-s)/\tau_{eq}} (4 \pi
          (t-s))^{-\frac{d}{2}}
      \EEA
      and only a dependence on the time difference $t-s$ is observed.
\end{itemize}
Comparing the values of non-equilibrium exponents derived from the autoresponse with those obtained from 
the autocorrelation reveals that we have at criticality the identities $a_1^{(\tau)} = b_1^{(\tau)}$
and $\lambda_{R_1}^{(\tau)} = \lambda_{C_1}^{(\tau)}$ in agreement with the general scaling arguments
given in the introduction.

It has been argued \cite{Hen01} that in out-of-equilibrium dynamics the scaling functions of response functions
can be derived from the space-time symmetries of the corresponding noise-less Langevin equation. This has been 
checked for a series of exactly solvable systems (see \cite{Godreche2002,Henk04} 
and references therein), among them the spherical model in the bulk system
\footnote{Note, however, that in critical systems with $z \neq2$, as for example in the critical two- and
three-dimensional Ising models, corrections to the prediction of local scale invariance (LSI) have been shown to
exist \cite{PleGam05}. But even in that case the scaling function coming from the theory of local scale invariance \cite{Hen02}
yields an excellent description of the numerically determined autoresponse function.}. 
In \cite{Plei04} local scale invariance was used to derive the following prediction for the surface autoresponse
function (with $t > s$):
\BEQ \label{gl:lsi_surf}
R_1^{LSI}(t,s) = r_0 \left( \frac{t}{s} \right)^{\zeta_2 - \zeta_1} ( t-s )^{\zeta_1 + \zeta_2}
\EEQ
where $\zeta_1$ and $\zeta_2$ are two exponents left undetermined by the theory, and $r_0$ is a non-universal normalization
constant. The values of $\zeta_1$ and $\zeta_2$ are fixed by comparing (\ref{gl:lsi_surf}) with the expected scaling behaviour (see Equations
(\ref{R1}) and (\ref{fc1})) yielding the result
\BEQ \label{gl:lsi_surf2}
R_1^{LSI}(t,s) = r_0 \, s^{-1-a_1^{(\tau)}} \left( \frac{t}{s} \right)^{-1-a_1^{(\tau)}} \left( \frac{t}{s} -1
\right)^{1+a_1^{(\tau)}-\lambda_{R_1}^{(\tau)}/z}.
\EEQ
This prediction is in complete agreement with our exact results for $T \leq T_c$ both for Dirichlet and
Neumann boundary conditions.

Finally, let us mention that it was shown in \cite{Hen94} that the dependence of the full response function 
(\ref{gl:response_res0}) on $r$ and $r'$ can be deduced from the space-time symmetries
in the case of Dirichlet boundary conditions.

\section{The fluctuation-dissipation ratio}
The fluctuation-dissipation ratio
\BEQ
\label{gl:diss_fluc}
X(t,s) := \frac{T \, R(t,s)}{\partial_s C(t,s)}
\EEQ
has been discussed extensively in recent years \cite{Cris03} as a possible way for attributing an effective
temperature to an out-of-equilibrium system. Of importance in the following is the fact that a characteristic
behaviour is expected for different physical situations. Thus in systems undergoing phase ordering one usually
expects $X(t,s)$ to approach 0, whereas a different behaviour is observed for non-equilibrium critical systems.
Indeed it has been realized \cite{Godreche2000,CalGam05} that in the latter case the non-vanishing limit value
\BEQ
\label{gl:diss_fluc2}
X^{\infty} := \lim_{s \rightarrow
\infty} \lim_{t \rightarrow \infty} X(t,s)
\EEQ
is a universal quantity whose value characterises the given dynamical universality class.

The concept of fluctuation-dissipation ratio, initially introduced for bulk systems,
has been generalized in \cite{Plei04} to systems with surfaces.
The surface fluctuation-dissipation ratio is thereby defined by
\BEQ
\label{gl:diss_fluc3}
X_1(t,s) := \frac{T \, R_1(t,s)}{\partial_s C_1(t,s)}
\EEQ
The asymptotic value $X_1^{\infty} := \lim_{s \rightarrow
\infty} \lim_{t \rightarrow \infty} X_1(t,s)$ is in fact a ratio of two amplitudes and
its value at the bulk critical point should be characteristic for a given surface universality class \cite{CalGam05}.
Recently, $X_1(t,s)$ has been determined numerically for the critical two- and three-dimensional 
semi-infinite Ising models \cite{Plei04}, with
asymptotic values $X_1^{\infty}$ differing from the values $X^{\infty}$ obtained in the corresponding
bulk systems. $X_1^{\infty}$ has also been computed within the Gaussian model \cite{CalGam05} 
yielding the value $\frac{1}{2}$. 

Having already determined the surface autocorrelation and autoresponse functions in the previous sections,
we are in the position to compute $X_1(t,s)$ also for the semi-infinite spherical model. In doing so
one immediately realizes that for Neumann boundary conditions $X_1^{(n)}(t,s)$ is identical to the
bulk quantity $X(t,s)$ obtained in \cite{Godreche2000}. This follows from the fact that $C_1$ and $R_1$ for
Neumann and periodic boundary conditions only differ by a numerical constant which drops out when the ratio
is formed. For this reason we give here only the results obtained for Dirichlet boundary conditions:
\begin{itemize}
  \item \underline{$T < T_C$:}
  \BEQ
  X_1^{(d)}(t,s)  = \frac{4 (8 \pi)^{-\frac{d}{2}}}{M_{eq}^2} s^{-\frac{d}{2}+1} 
  \left( \frac{\frac{t}{s}+1}{\frac{t}{s}-1} \right)^{\frac{d}{2}+1} 
  \frac{\frac{t}{s} + 1}{(4+d)\frac{t}{s} -d }  
  \EEQ
  yielding the limit value $X_1^{\infty} = 0$ as expected for ferromagnetic systems quenched below their critical point.
  \item \underline{$T = T_C$ and $2 < d < 4$:}
  In this case a straightforward but somewhat tedious calculation yields
  \BEQ
  \hspace{-2.0truecm}
  X_1^{(d)}(t,s) = \frac{d (d -2) (\frac{t}{s} + 1)^3}{(d^2-16) + 
  (32 + d(3 d - 8))\frac{t}{s} + (d-4)(4+3d) (\frac{t}{s})^2 + d^2 (\frac{t}{s})^3}
  \EEQ
  with the limit value $X_1^{\infty} = \frac{d-2}{d}$.
  \item \underline{$T = T_C$ and $ d > 4$:}
  Here we obtain the same expressions as for Neumann and periodic boundary conditions:
  \BEQ
  X_1^{(d)}(t,s) = \frac{1}{1 +
      \left(\frac{\frac{t}{s}-1}{\frac{t}{s}+1}\right)^{\frac{d}{2}+1}}
  \EEQ
  with $X_1^{\infty} = \frac{1}{2}$.
\end{itemize}
In all cases the limit value turns out to be independent of the boundary condition, which is the reason why we
have dropped the superscript $(d)$. We therefore conclude that in the semi-infinite spherical model $X_1^{\infty}$ equals
$X^{\infty}$ independently of the chosen boundary condition.

\section{Conclusion}
In this paper we have extended the study of out-of-equilibrium dynamical properties of 
the kinetic spherical model to the semi-infinite geometry with both Dirichlet and Neumann boundary conditions.
The exact computation of two-time surface quantities (like the autocorrelation and the autoresponse functions)
reveal that dynamical scaling is also observed close to a surface for quenches to temperatures below or equal
to the critical temperature. Whereas for Neumann boundary conditions we find that the values of the non-equilibrium
exponents and the scaling functions (up to a numerical factor) are identical to the bulk ones, the situation for
Dirichlet boundary conditions is more interesting. Indeed at the critical point we find that out-of-equilibrium
dynamics is governed by universal exponents whose values differ from those of the corresponding bulk exponents.
The values of these surface exponents are in complete agreement with predictions coming from general scaling
considerations \cite{Plei04,CalGam05}. Similarly, surface scaling functions are also found 
to differ from the bulk scaling functions. Interestingly, we find that in the ordered low-temperature phase
the autocorrelation function scales in the ageing regime as
\BEQ
C_1(t,s) = s \, f_{C_1}(t/s),
\EEQ
in strong contrast to the well-known bulk behaviour \cite{Bray94}
\BEQ 
C(t,s)  = f_C(t/s).
\EEQ
As this paper is the first study of ageing phenomena close to a surface in an ordered phase, it is an open
und interesting question whether this is only a special feature of the kinetic spherical model or whether
this is a general property of semi-infinite systems undergoing phase ordering. We intend to come back to this
problem in the near future.

\appsection{A}{Eigenvalues and eigenvectors of the interaction matrix $\mathbf{Q}_{\Lambda}^{(\tau)}$}
The eigenvalues $\mu_{\Lambda}^{(\tau)}(\vec{k})$
of the interaction matrix $\mathbf{Q}_{\Lambda}^{(\tau)}$ depend on the
boundary conditions and are given by \cite{Brankov}
\BEQ
\mu_{\Lambda}^{(\tau)}(\vec{k}) = 2 \sum_{\nu=1}^d \cos
\left(\varphi^{(\tau_{\nu})}_{L_{\nu}} (k_{\nu})\right), \quad \vec{k} \in
\Lambda
\EEQ
where the functions $\varphi_L^{(\tau)}$ are
\BEA
\label{gl:phi1}
\varphi_L^{(p)}(k) &= \frac{2 \pi k}{L} & \qquad
\mbox{(periodic)} \\
\label{gl:phi2}
\varphi_L^{(d)}(k) &= \frac{\pi k}{L+1} & \qquad
\mbox{(Dirichlet)} \\
\label{gl:phi3}
\varphi_L^{(n)}(k) &= \frac{\pi (k-1)}{L} & \qquad
\mbox{(Neumann)} 
\EEA
The (orthonormal) eigenvectors are given by the expressions
\BEQ
u_{\Lambda}^{(\tau)}(\vec{r},\vec{k}) = \prod_{\nu = 1}^d
u_{L_{\nu}}^{(\tau_{\nu})}(r_{\nu},k_{\nu})
\EEQ
with
\BEA
\label{gl:eig1}
u_L^{(p)}(r,k) &=& L^{-\frac{1}{2}} \exp(-\II r
\varphi_L^{(p)}(k)) \\
\label{gl:eig2}
u_L^{(d)}(r,k) &=& [2/(L+1)]^{\frac{1}{2}} \sin(r
\varphi_L^{(1)}(k)) \\
\label{gl:eig3}
u_L^{(n)}(r,k) &=& \left\{ \begin{array}{ll}
                           L^{-\frac{1}{2}} & k = 1 \\
			   (2/L)^{\frac{1}{2}} \cos(
			   (r-\frac{1}{2})
			   \varphi_L^{(0)}(k)) & k =
			   2,\ldots,L
\end{array} \right.
\EEA

\appsection{B}{Asymptotic behaviour of $g$}
In this Appendix we recall the asymptotic behaviour \cite{Godreche2000} of the function
\BEQ
g(t) = \hat{f}(t) + 2 T \int_0^t \D u \hat{f}(t-u)g(u)
\EEQ
with
\BEQ
\label{gl:const_function2}
\hat{f}(t) := \int_{-\pi}^{\pi} \frac{ \D^{d-1} \vec{q}}{(2
\pi)^{d-1}} \int_{-\pi}^{\pi} \frac{\D k}{2 \pi}
e^{-2 \omega(k,\vec{q}) t} = (e^{-4 t} I_0(4 t))^d.
\EEQ
Defining first the constants
\BEA
A_k &:=& \int \frac{\D^{d-1} \vec{q}}{(2 \pi)^{d-1}}
\int_{-\pi}^{\pi} \frac{\D k}{2 \pi} \frac{1}{(2 \omega(k,\vec{q}))^k}
\EEA
the critical temperature can be expressed as
\BEA
T_C & := & \frac{1}{2 A_1}.
\EEA
The asymptotic behaviour of $g$ for long times, denoted $g_{age}$, is then given by
\begin{itemize}
  \item \underline{$T>T_C$:}
  \BEQ
  g_{age}(t) \stackrel{t \rightarrow \infty}{\sim}e^{t/\tau_{eq}}
   \EEQ
  where the exact expression for the time scale
  $\tau_{eq}$ can be found in \cite{Godreche2000},
  \item \underline{$T<T_C$:}
  \BEQ
  g_{age}(t) \stackrel{t \rightarrow \infty}{\sim}
  \frac{(8 \pi t)^{-\frac{d}{2}}}{M_{eq}^2} \quad
  \mbox{with} \quad M_{eq}^2 = 1 - \frac{T}{T_C},
  \EEQ
  \item \underline{$T = T_C$ and $2 < d < 4$:}
  \BEQ
  g_{age}(t) \stackrel{t \rightarrow \infty}{\sim}(d-2)(8 \pi)^{\frac{d}{2}-1} \sin
   \left(\frac{(d-2)\pi}{2} \right)
  \frac{t^{-(2-\frac{d}{2})}}{T_C^2},
  \EEQ
  \item \underline{$T = T_C$ and $d > 4$:}
  \BEQ
  g_{age}(t) \stackrel{t \rightarrow
  \infty}{\rightarrow} \frac{1}{4 A_2 T_C^2}.\\
  \EEQ
  \end{itemize}

\noindent
{\bf Acknowledgements:}\\~\\
We acknowledge the support by the Deutsche Forschungsgemeinschaft
through grant no. PL 323/2.\\~\


\end{document}